\documentclass[nofootinbib,12pt,onecolumn]{revtex4}
\usepackage[a4paper,includeheadfoot,margin=2.7cm]{geometry}
\usepackage{graphicx}
\usepackage{subfig}
\usepackage{amssymb}
\usepackage{amsmath}
\usepackage{bm}
\usepackage{slashbox}
\begin{document}

\title{To conserve, or not to conserve: A review of nonconservative theories of gravity}
	
\author{Hermano Velten$^1$}\email{hermano.velten@ufop.edu.br}
\author{Thiago R. P. Caram\^es$^2$}\email{thiago.carames@ufla.br}

\affiliation{$^1$ Departamento de F\'isica, Universidade Federal de Ouro Preto (UFOP), Ouro Preto-MG, Brazil}
\vspace{1cm}

\affiliation{$^2$ Departamento de F\'isica, Universidade Federal de Lavras (UFLA), Lavras-MG, Brazil}

%\author{Thiago~R.~P.~Caram\^es,$^a$\footnote{thiago.carames@ufes.com}~~J\'ulio~C.~Fabris,$^a$\footnote{fabris@pq.cnpq.br}\\ E.~R.~Bezerra de Mello,$^b$\footnote{emello@fisica.ufpb.br}~~H.~Belich,$^a$\footnote{humberto.belich@ufes.br}\vspace{0.5cm}\\
%$a$ Universidade Federal do Esp\'{\i}rito Santo (UFES), Av. Fernando Ferrari-514, 29075-910, Vit\'oria, ES - Brazil \\
%$b$ Departamento de F\'isica, Universidade Federal da Para\'iba
%58.059-970, Caixa Postal 5.008, Jo\~ao Pessoa, PB - Brazil \\}

\begin{abstract}

Apart from the familiar structure firmly-rooted in the general relativistic field equations where the energy--momentum tensor has a null divergence i.e., it conserves, there exists a considerable number of extended theories of gravity allowing departures from the usual conservative framework. Many of these theories became popular in the last few years, aiming to describe the phenomenology behind dark matter and dark energy. However, within these scenarios, it is common to see attempts to preserve the conservative property of the energy--momentum tensor. Most of the time, it is done by means of some additional constraint that ensures the validity of the standard conservation law, as long as this option is available in the theory. However, if no such extra constraint is available, the theory will inevitably carry a non-trivial conservation law as part of its structure. In this work, we review some of such proposals discussing the theoretical construction leading to the non-conservation of the energy--momentum tensor.
 
\textbf{Key-words}: general relativity; cosmology; extended theories of gravity
%PACS numbers: $04.50.+h$, $04.20.-q$
\end{abstract}

\maketitle

\section{Introduction}

The principle of matter-energy conservation is one of the main pillars of General Relativity (GR). Its importance when formulating a generally covariant gravitational theory is a matter of intense discussion since the first few years of GR. Indeed, the seminal work by E. Noether has its origins in the debate between F. Klein, E. Noether, D. Hilbert and A. Einstein about the mathematical relevance of energy conservation (see Ref.~\cite{Brading:2005ina} for historical details). Throughout the 20th century, the features of conservation laws in GR have been frequently discussed in the literature~\cite{Goldberg:1958zz,Komar:1958wp,Bergmann:1958zz,Bondi1990,Bak:1993us}.

Within a general relativistic based description of gravity, the covariant conservation law obeyed by ordinary matter is straightforwardly obtained when one applies the contracted Bianchi identities to the Einstein equations, which provides the well-known null covariant derivative of the energy--momentum tensor for the respective gravitating system. However, there is much more than a mere mathematical result in this important aspect of General Relativity (GR). It reveals two paramount features of the Einstein's gravity: the invariance under diffeomorphism and the minimal matter--curvature coupling. The former aspect means, in other words, that GR is a coordinate invariant theory, whereas the latter reflects the clear separation between the geometrical and matter sectors seen in the effective action of the theory. 

Thus, one expects that any extended gravity theory evading some of these properties shall lead to a different conservation condition to be obeyed by a given energy--momentum tensor. It is possible, however, to have a deviation from the usual conservation law by imposing it by hand. A famous example is the Rastall gravity~\cite{Rastall:1973nw}. In addition, like Brans--Dicke theory, popular in some scalar tensor theories, non-conservation can also be achieved if one works within the Einstein frame~\cite{BD,Magnano,Faraoni}, where the dilaton comes up as part of the matter sector. Many other works can be found in the literature dedicated to analyzing the arising of non-conservation within the context of alternative theories of gravity. In this review, we shall revisit some of them. 

The breaking of diffeomorphism may be verified in the gravity sector of the so-called Standard Model Extension (SME), in general accompanied also by a local violation of Lorentz symmetry~\cite{bluhm}. As discussed in such a reference, the breaking is caused by the presence of a background field, which can either be endowed with dynamics or not. When this field has a dynamical character, the standard conservation of the energy--momentum tensor is naturally obeyed.  On the other hand, when this field has no dynamics, the breaking it induces is denoted ``explicit'' and leads to a deviation from the usual conservation law. In models where gravity is thought as an emergent phenomenon, i.e., a low energy manifestation of a fundamental higher energy theory where a background dependence shows up, diffeomorphism breaking is also verified, as discussed in~\cite{Anber:2009qp}. In that work, as expected, the vanishing of the covariant divergence of a given energy--momentum tensor is not automatically satisfied. Actually, the authors had to impose it by assuming an additional constraint in the model.  

Another nonconservative gravity that has attracted recent attention is the Lazo's theory, in which the Lagrangian density carries a dependence on the action itself~\cite{Lazo:2017udy}. This theory consists of a covariant version of the Herglotz variational problem, which by its turn was an attempt to incorporate dissipative effects into the classical mechanics via a variational principle~\cite{herglotz}. In Lazo's approach, the non-conservation of energy--momentum is caused by the presence of a background four-vector that introduces into the theory a preferred direction, thus breaking the diffeomorphism invariance.  

As mentioned above, models where matter and gravity are non-minimally coupled constitute another realm where a departure of the usual conservation law is verified. In~\cite{koivisto}, the author discusses how this conservation condition should look for a wide class of such modified gravity theories, both in the metric and Palatini formulations. Considering a family of models whose action carries both non-minimal coupled terms and arbitrary functions of the scalar curvature, he finds expressions for the modified conservation law for both of the variational formalisms. In addition, he shows that it is possible to generalize the Bianchi identity so that the usual conservation law is ensured. This result arises thanks to a specific choice of boundary conditions made during the process of extremization of the action under a given infinitesimal active coordinate transformation. In~\cite{tian}, the authors obtain the energy--momentum conservation for an even wider class of theories of gravity, where the geometric dependence of the non-minimal coupling function is not restricted to scalar curvature, as it may also depend on the square of the Ricci and Riemann tensors. Furthermore, they also consider another family of non-minimal coupled theories where Lagrangian density has an arbitrary dependence on multiple curvature invariants. For both cases, they obtained the extended conservation law, generalizing previous results.  

An interesting case where such a non-minimal interaction between matter and curvature is also admissible, having expected consequences for the covariant conservation law, is the family of the so-called $f(R,T)$ theories \cite{Harko:2011kv}. There are, however, specific functional forms for $f(R,T)$ in which the standard conservation can be preserved\cite{alvarenga}. %MDPI: is the bold necessary.
On the other hand, it is worth mentioning that, in this conservative subclass, there is no mixing involving the both dependencies on $R$ and $T$. In other words, the density Lagrangian $f(R,T)$ admits the particular form $f(R,T)=f_1(R)+f_2(T)$. In this vein, it is possible to use the purely $T$-dependent term as part of a redefinition of the matter sector in a minimally coupled gravity~\cite{fisher}. This aspect helps us to illustrate the close relation between the matter/curvature coupling and the conservation law to be obeyed by an energy--momentum tensor.

Apart from proposals of modified gravitational theories, it is worth mentioning that the steady state cosmological model, proposed by T. Gold, H. Bondi, and F. Hoyle (see~\cite{Hoyle,Bondi}), proposes that the universe expands eternally, with continual creation of matter assuring a constant density of mass. It became clear that predictions of the steady state model were not compatible with new observational data that supported the Big Bang cosmology. The concept of matter creation is still present in modern phenomenological models to mimic a possible interaction between dark matter and dark energy~\cite{Fritzsch:2012qc,Pigozzo:2015swa}. Even in the context of modified gravity theories, the cosmological particle creation process has been investigated~\cite{Koutsoumbas:2013boa,Ema:2015oaa,Capozziello:2016pyv,Yu:2018qzl}.   

In the next section, we review the notion of energy--momentum conservation tensor in General Relativity. The interpretation of conserved quantities in a gravitational field background is a very subtle issue and deserves a proper discussion. In the subsequent sections, we present specific nonconservative theories. At the end of this work, we revisit the notion of energy conditions in modified gravity theories (Section \ref{sec:energy}) and then present our final considerations.

\section{The Conservative Landscape}
\label{sec:2}
\subsection{From Special to General Relativity}

The conservation principles are one of the most interesting aspects of physics. They help us to make predictions about the evolution of a given physical system, ensuring that, despite the change, it undergoes certain aspects present in it that shall remain the same. Already in our high school physics and early undergraduate classes, we made contact with such an important property and learned that, under certain circumstances, the energy, linear, and angular momentum of physical systems are preserved (see~\cite{Maudlin:2019bje} for a recent discussion). 

It is also usual to associate the notion of conservation with ideal physical situations in which dissipative mechanisms do not take place. Indeed, one of the main pillars of physics is the  Hamilton’s principle of stationary action used to derive equations of motion for many conservative systems of varying degrees of complexity. It is worth noting that only recently  an extension of the Hamilton's principle to nonconservative classical systems has been developed \cite{Galley:2012hx}.

From a strictly, but crude, mathematical standpoint, the notion of conservation is intrinsically related to the way one performs derivatives of physical quantities. Our usual concept of conservation has its foundations in simple laboratory experiments in which classical systems e.g., hydrodynamics experiments, are tested. Starting with a flat spacetime metric with signature $\eta_{\mu\nu}=(-1,+1,+1,+1)$, one defines the derivative of a scalar quantity $\varphi$ simply as $\partial \varphi / \partial x \equiv \varphi_{,x}$ or, in a multi-dimensional spacetime with coordinates denoted by index $\mu$, i.e., $\partial \varphi / \partial x^{\mu} \equiv \varphi_{,\mu}$. Here, the symbol comma $``,``$ refers to an ordinary derivative. However, the formulation of currents and the energy conservation has revealed much more intricacies than that~\cite{Forger:2003ut}. 

The tensorial formalism is a more generic structure to represent fluid quantities. In any spacetime, the energy--momentum tensor $T^{\mu\nu}$ can be decomposed in its rest frame components such that $T^{00}=\rho=$ energy density; $T^{0i}$ represents the internal heat-conduction; $T^{i0}$ corresponds to the momentum transferred in the internal energy flux process and $T^{ij}$ is the momentum flux. This tensor is symmetric so that $T^{\mu\nu}=T^{\nu\mu}$. For a fluid element occupying a non expanding volume subjected to energy/particle flowing across its surface, the conservation of energy is stated by $T^{0 \mu}_{\quad, \mu}=0$ while momentum conservation by $T^{i \mu}_{\quad, \mu}=0$ where $i$ refers to the momentum component under analysis. This implies in the general conservation law
\begin{equation}\label{ConsFlat}
    T^{\mu\nu}_{\quad,\mu}=0.
\end{equation}

Along the fluid flow, one also defines the particle quadriflux with components $N^{0}=c \times$ particle number density and $N^i=$ particle flux. The macroscopic description of relativistic fluids demands the introduction of the four-velocity $u^{\mu}$ for which by convention one has $u_{\mu}=\eta_{\mu\nu} u^{\nu}$ with $u_0=-1$ and $u_i=0$. Therefore, $N^{\mu}=n u^{\mu}$. The particle flux conservation is then expressed by the law
\begin{equation}
    N^{\mu}_{\quad,\mu}=(n u^{\mu})_{,\mu}=0.
\end{equation}

Apart from vacuum solutions e.g., black holes, in which the intrinsic gravitational aspects are studied, it is obvious that the universe is not empty. It is therefore mandatory to set up an energy--momentum tensor for relativistic fluids. The simplest possible configuration, and widely used as the standard starting point in the study of relativistic fluids, is to consider the so-called perfect fluids. They are basically non-viscous fluid configurations obeying the structure
\begin{equation}\label{EM}
    T^{\mu\nu}=(\rho + p)u^{\mu}u^{\nu}+p \eta^{\mu\nu}=\rho u^{\mu}u^{\nu}+p h^{\mu\nu},
\end{equation}
where $\rho$ is the energy density and $p$ the fluid pressure. The second equality of (\ref{EM}) states that the pure pressure contribution is associated with the symmetric projection tensor $h^{\mu\nu}=u^{\mu}u^{\nu}+\eta^{\mu\nu}$.

In the context of a covariant gravitational theory as, e.g., GR, the manifestation of the gravitational interaction is seen as an effect of the curvature of the spacetime. Trajectories, flows, and the variation rates of physical quantities should obey new rules that take into account curvature. The mathematical mechanism used to address this issue is the replacement of the flat spacetime metric $\eta^{\mu\nu}$ by the curved one $g^{\mu\nu}$. The metric $g^{\mu\nu}$ is adapted to the physical problem one wants to study and is written in such a way that it describes the geometry of the curved spacetime. Now, the equivalence principle implies that the conservation law (\ref{ConsFlat}) is replaced by its version in a curved spacetime 
\begin{equation}\label{ConsCurve}
    T^{\mu\nu}_{\quad ; \mu}=T^{\mu\nu}_{\quad , \mu} + T^{\alpha \nu}\Gamma^{\mu}_{\alpha \mu}+T^{\mu\alpha}\Gamma^{\nu}_{\alpha\mu}=0,
\end{equation}
where the symbol ``;'' means covariant derivative. The additional contributions on the right-hand side brings the so-called affine connection, which, for a Riemannian manifold, coincide with the Christoffel symbols, defined as follows:
\begin{equation}
   \Gamma^{\alpha}_{\mu\nu} =\frac{g^{\alpha\beta}}{2}\left(g_{\beta\mu,\nu}+g_{\beta\nu,\mu}-g_{\mu\nu,\beta}\right).
\end{equation}

There are four different equations within (\ref{ConsCurve}) since $\nu=0,1,2,3$. The $\nu=0$ equation denotes conservation of energy while, for $\nu=i=1,2,3$, one has the conservation of the $i^{th}$ component of the momentum.

According to our discussion so far, Equation (\ref{ConsCurve}) has been introduced as an extension of the flat spacetime conservation to curved geometries. The gravitational interaction is not implicitly stated at this stage. However, we know that matter curves spacetime via gravity. Then, prior to the appropriate introduction of the gravitational interaction in our discussion, let us continue to describe generic curved manifolds via the definition of the Riemann curvature tensor
\begin{equation}
    R^{\alpha}_{\;\;\beta \mu\nu}=\Gamma^{\alpha}_{\beta\nu,\mu}-\Gamma^{\alpha}_{\beta\mu,\nu}+\Gamma^{\alpha}_{\sigma\mu}\Gamma^{\sigma}_{\beta\nu}-\Gamma^{\alpha}_{\sigma\nu}\Gamma^{\sigma}_{\beta\mu}.
\end{equation}

Using the fact that second derivatives of the metric tensor are non-vanishing quantities and that partial derivatives commute, the following identity takes place: 
\begin{equation}
    R_{\alpha\beta\mu\nu}+R_{\alpha\nu\beta\mu}+R_{\alpha\mu\nu\beta}=0.
\end{equation}

Similarly, one can find symmetry properties of the Riemann tensor e.g., $R_{\alpha\beta\mu\nu}=-R_{\beta\alpha\mu\nu};\,\,  R_{\alpha\beta\mu\nu}=-R_{\alpha\beta\nu\mu}; \,\,R_{\alpha\beta\mu\nu}=R_{\mu\nu\alpha\beta}$. Finally, with such results, one can find the desired results for our discussion, the so-called Bianchi identities
\begin{equation}\label{Bianchi}
    R_{\alpha\beta\mu\nu;\lambda}+R_{\beta\lambda\mu\nu;\alpha}+R_{\lambda\alpha\mu\nu;\beta}=0.
\end{equation}

Now, we can show the consequence of this identity to the conservation of $T^{\mu\nu}$. By contracting Ref. (\ref{Bianchi}) twice, firstly with $g^{\alpha \mu}$, then, with $g^{\beta\nu}$ and using the symmetry properties of the metric tensor, the Bianchi identity becomes 
\begin{equation}\label{Bianchi2}
    (2R^{\mu}_{\lambda}-\delta^{\mu}_{\lambda} R)_{; \mu}=0,
\end{equation}
where the Ricci tensor $R_{\alpha\beta}=R^{\mu}_{\;\;\alpha\mu\beta}=R_{\beta\alpha}$ and the Ricci scalar $R=g^{\mu\nu}R_{\mu\nu}$ have been defined. In principle, there is nothing special with Equation (\ref{Bianchi2}). Let us analyze an important consequence of (\ref{Bianchi2}). As it is well known, the GR field equations may be derived from a variational principle. The starting point of this procedure is the total action below 
\begin{equation}
\label{gravAc}    
S=\frac{1}{2\kappa} \int d^4 x \sqrt{-g} R+\int d^4x \sqrt{-g}{\cal L}_{m}.   
\end{equation}

The first term on the right-hand side is the so-called Einstein--Hilbert action, defined via the Lagrangian $\mathcal{L}_{EH}=R$, whilst the second one is the matter action defined as usual, in terms of the Lagrangian density associated with the matter fields, ${\cal L}_{m}$. The energy--momentum tensor of arbitrary matter configurations is defined in terms of ${\cal L}_{m}$ in the following way:
\begin{equation}
\label{emt}
T_{\mu\nu}= \frac{2}{\sqrt{-g}}\frac{\delta \left(\sqrt{-g} {\cal L}_m\right)}{\delta g^{\mu\nu}}.
%\label{defTmunu}
\end{equation}

Taking (\ref{emt}) into account, the variation of the action (\ref{gravAc}) gives
%Let us start from
%\begin{equation}
%    \delta \int d\Omega \sqrt{-g} \mathcal{L}=0
%\end{equation}
%where the invariant volume element reads $d\Omega %\sqrt{-g}$ and the Lagrangian density is $\mathcal{L}$. %The choice for the latter leading to the gravitational %sector of GR field equations is $\mathcal{L}=R$ yielding %to
%\begin{equation}
%\int d\Omega \sqrt{-g} \delta %g^{\mu\nu}\left(R_{\mu\nu}-\frac{1}{2}R %g_{\mu\nu}\right)=0
%\end{equation}
%If we add to $\mathcal{L}$ the matter sector i.e., now there is an extra matter Lagrangian density to the total one finds the widely known GR field equation
\begin{equation}\label{EinsteinEq}
    R_{\mu\nu}-\frac{1}{2}g_{\mu\nu}R={\kappa}T_{\mu\nu},
\end{equation}
where one immediately recognizes the left-hand side of this equation, also known as the Einstein tensor
\begin{equation}
G_{\mu\nu}=R_{\mu\nu}-\frac{1}{2}g_{\mu\nu} R,
\end{equation}
with the quantity appearing in (\ref{Bianchi2}). Therefore, the covariant derivative in (\ref{Bianchi2}) should also apply to the right-hand side of (\ref{EinsteinEq}) implying conservation of $T_{\mu\nu}$. In the equation above, $\kappa\equiv 8\pi G$ is the gravitational coupling constant by assuming that we are working with $c=1$ units.

Alternatively, in the presence of a cosmological constant, the Einstein--Hilbert Lagrangian would be redefined as ${\cal L}_{EH}\rightarrow {\cal L}_{EH}-2\Lambda$. The resulting field equations in this case are given by
\begin{equation}\label{EinsteinEqLambda}
    G_{\mu\nu}+\Lambda g_{\mu\nu}={\kappa}T_{\mu\nu}.
\end{equation}

The above equation also has vanishing covariant divergence, and it is the only field equation of the generic type $F_{\mu\nu}(g_{\alpha\beta},g_{\alpha\beta,\delta},g_{\alpha\beta,\delta\gamma})=T_{\mu\nu}$, where $F_{\mu\nu}$ is a tensor functional, derivable from an action principle in which the gravitational Lagrangian density is a scalar invariant of the metric~\cite{Lovelock:1971yv}.

\subsection{Diffeomorphism Invariance}

A remarkable aspect of GR arises when we consider the invariance of the theory under diffeomorphism. Consider an infinitesimal active transformation generated by a given vector field $V^{\mu}$. This corresponds to the following mapping
\begin{equation}
\label{active}    
x^{\mu}\rightarrow x^{\mu}+V^{\mu}.    
\end{equation}

It is well known that such a transformation allows us to introduce a derivative operator which provides the rate of change of a given tensor along the integral curves of $V^{\mu}$. This operator is the so-called Lie derivative, denoted as ${\cal L}_{V}$ (where the index refers to $V^{\mu}$). In any GR textbook, the reader can find a detailed discussion on how such an operator acts on arbitrary-rank tensors~\cite{carrol,wald,dinverno}. An interesting relation shows up when one applies such operator on the metric tensor. In this case, we have 
\begin{equation}
\label{Lvg}
{\cal L}_{V}g_{\mu\nu}=2\nabla_{(\mu}V_{\nu)}.    
\end{equation}

In order to proceed with our purpose, let us rewrite the total action (\ref{gravAc}) as follows:
\begin{equation}
\label{gravAc1}
S=\frac{1}{2\kappa}S_{EH}[g_{\mu\nu}]+S_{m}[g_{\mu\nu},\psi^{i}].     
\end{equation}

The theory is diffeomorphism-invariant if such a transformation implies in a variation of the action so that $\delta S=0$. Thus, let us assume this property a priori and examine its consequences. By varying (\ref{gravAc1}) and making it equal to zero, we have
\begin{equation}
\label{delS}
\frac{1}{2\kappa}\int d^4x \frac{\delta\left(\sqrt{-g}{\cal L}_{EH}\right)}{\delta g_{\mu\nu}}\delta g_{\mu\nu}+\int d^4x \frac{\delta \left(\sqrt{-g} {\cal L}_{m}\right)}{\delta g_{\mu\nu}}\delta g_{\mu\nu}+\int d^4x \sqrt{-g} \frac{\delta S_{m}}{\delta \psi^{i}}\delta \psi^{i}=0 \end{equation}

For a diffeomorphism generated by $V^{\mu}$, the variation of the metric is simply its Lie derivative along $V^{\mu}$, which is given by (\ref{Lvg}). We can check that the Einstein--Hilbert action is itself invariant under diffeomorphism. To verify this, notice that the first term of (\ref{delS}) corresponds to
\begin{eqnarray}
\label{EHvar}
\delta S_{EH}=\int d^4x \frac{\delta\left(\sqrt{-g}{\cal L}_{EH}\right)}{\delta g_{\mu\nu}}\delta g_{\mu\nu}&=&\int d^4x \sqrt{-g} G^{\mu\nu}\delta g_{\mu\nu}\nonumber\\ &=&\int d^4x \sqrt{-g} G^{\mu\nu}{\cal L}_{V}g_{\mu\nu}\nonumber \\ &=&2\int d^4x \sqrt{-g} G^{\mu\nu}\nabla_{\mu}V_{\nu}. 
\end{eqnarray}

If we now integrate (\ref{EHvar}) by parts and use the covariant form of the Gauss law, we find
\begin{eqnarray}
\label{EHvar1}
\delta S_{EH}&=&\int d^4x \sqrt{-g}\left(\nabla_{\mu} G^{\mu\nu}\right)V_{\nu},
\end{eqnarray}
where we get rid of the surface integral, as it is assumed that the vector $V^{\mu}$ vanishes on the boundary, although $V^{\mu}$ is arbitrary within the volume enclosed by such a boundary. Thus, for a generic $V^{\mu}$, (\ref{EHvar1}) tells us that the invariance of diffeomorphism of the Einstein--Hilbert action, $\delta S_{EH}=0$, is ensured by $\nabla_{\mu} G^{\mu\nu}=0$, which is the contracted Bianchi identity, previously presented in (\ref{Bianchi2}).

It is implicit that the fields $\psi^{i}$ satisfy the matter equations of motion, which leads to the third term vanishing. Thus, the remaining term has necessarily to obey
\begin{equation}
\label{lm}
\int d^4x \frac{\delta \left(\sqrt{-g} {\cal L}_{m}\right)}{\delta g_{\mu\nu}}\delta g_{\mu\nu}=0.
\end{equation}

Using the definition (\ref{emt}) and (\ref{Lvg}) in (\ref{lm}), one finds
\begin{equation}
\label{lm1}
\int d^4x \frac{\sqrt{-g}}{2}T^{\mu\nu}{\cal L}_{V}g_{\mu\nu}=0.    
\end{equation}

Analogous to the previous case, we can repeat here the procedure described by (\ref{EHvar}) and (\ref{EHvar1}). This shall lead (\ref{lm1}) to
\begin{equation}
\label{lm1}
\int d^4x \sqrt{-g}\left(\nabla_{\mu}T^{\mu\nu}\right)V_{\nu}=0.    
\end{equation}

Given an arbitrary $V^{\mu}$, (\ref{lm1}) implies in the covariant conservation of $T_{\mu\nu}$, namely (\ref{ConsCurve}) 
\begin{equation}
%\label{consT}
\nabla_{\mu}T^{\mu\nu}=0.
\end{equation}

Now, we understand how the standard energy--momentum conservation emerges within GR as a product of an essential property of the theory, i.e., the diffeomorphism invariance. In the first few years after General Relativity was formulated, an intense debate existed on whether or not energy conservation was a true mathematical identity of the theory~\cite{Brading:2005ina}. Finally, in \cite{noether1,noether2}, E. Noether showed that the conservation of energy, linear, and angular momentum of physical systems, as well as other physical quantities, are justified by first principles. The Noether’s claim is that the conservation of a given quantity follows from a specific symmetry obeyed by the action. In this vein, the time translation invariance leads to the energy conservation, whereas the position translation makes the linear momentum to be conserved, as it happens to the angular momentum when the action exhibits invariance under space rotations. This feature reveals a universal aspect of conservative systems in nature, and its validation is evoked in all physical domains ranging from the quantum world to the cosmos. There is also a recent attempt to provide a unified view on the conservation laws in gravitational systems~\cite{Obukhov:2014nja}.

The above discussion can not be seen as an argument to refute any alternative to equation (\ref{EinsteinEqLambda}) as long as $g_{\mu\nu}$ is the only field variable sourced by $T_{\mu\nu}$. In the next sections, we are going to show some remarkable examples in the literature.

\subsection{The Meaning of the Term ``Energy-Momentum Conservation'' in the Presence of a Gravitational Field}
Although we refer to Equation (\ref{ConsCurve}) in most of this work as a conservation law for the energy--momentum tensor, led mainly by the common usage of the term, it is important to emphasize that, roughly speaking, this classification may be misleading and not strictly correct if we think of how to extract in practice from $T_{\mu\nu}$ the physical quantities that will follow conservation laws, namely, the energy and momentum. This discussion is made by taking different paths in the various GR textbooks. As it is shown in~\cite{landau} in the Minkowski spacetime, the quantity below \footnote{Although we consider $c=1$ throughout this work, particularly in this section we show $c$ explicitly, in order to match the Ref.~\cite{landau} that is used in the present discussion.}
\begin{equation}
\label{Pi}    
P^{\mu}=\frac{1}{c} \int T^{\mu\nu} dS_{\nu}   
\end{equation}
is a conserved quantity identified with the $4-$momentum of the system. The integration is taken on the hypersurface $S$ which contains all the three-dimensional space. It is well known that the conservation of $P^{\mu}$ can be expressed in terms of the null divergence of $T^{\mu\nu}$:
\begin{equation}
\label{divT}
\partial_{\mu}T^{\mu\nu}=0,
\end{equation}
which makes it clear why the statement of (\ref{divT}) as a conservation law is totally consistent. 

It is natural to think of the extension of (\ref{Pi}) in a curved spacetime as follows:
\begin{equation}
\label{4momentum}
P^{\mu}=\frac{1}{c}\int_{S} \sqrt{-g}T^{\mu\nu}dS_{\nu}.
\end{equation}

 We may wonder if does lead to conservation of energy and momentum in a curved manifold as (\ref{Pi}) does in the Minkowski spacetime. If this is true, the integral (\ref{4momentum}) would be conserved if the following condition applied:
\begin{equation}
\label{consC}    
\frac{\partial (\sqrt{-g}T^{\mu}_{\;\;\nu})}{\partial x^{\mu}}=0.    
\end{equation}

However, a mismatch is verified when we rewrite the equation (\ref{ConsCurve}) in the form below
\begin{equation}
\label{ConsCurve1}    
 T^{\mu}_{\;\;\nu\; ; \mu}=\frac{1}{\sqrt{-g}}\frac{\partial (\sqrt{-g}T^{\mu}_{\;\;\nu})}{\partial x^{\mu}}-\frac{1}{2}\frac{\partial g_{\mu\lambda}}{\partial x^{\nu}}T^{\mu\lambda}=0.
\end{equation}

The above expression is different from (\ref{consC}), unless we are in a particular coordinate system, $x^{*}$, where $\partial g_{\mu\lambda}(x^{*}) / \partial x^{\nu}=0$ holds. This result points to the need of reformulating the relation (\ref{4momentum}) in order to properly describe conservation of the energy and momentum in a curved spacetime. In this vein, Landau and Lifshitz call our attention to the fact that, although Equation (\ref{ConsCurve}) indeed expresses a local covariant conservation of $T^{\mu\nu}$, it is not true globally, since the gravitational field itself carries energy, whose contribution is missing in (\ref{4momentum}). This point is also illustrated by S. Weinberg in~\cite{weinberg}, by comparing the gravity with the electromagnetism. While in the latter case the electromagnetic field itself does carry charge, in the former, the gravitational field indeed carries energy and momentum that works as a source for gravity. This explains partially both the well-known linear nature of the Maxwell electromagnetic theory and the fully nonlinear character of GR. 
Thus, in~\cite{landau}, the author handles this issue by including into the model the contribution of the gravitational field. Such a contribution comes up in the form a pseudo-tensor $t^{\mu\nu}$, which is constructed with the aid of the geometric terms present in the Einstein equations. Thereby, they show that, instead of Equation (\ref{4momentum}), we shall have the following relation for the $4$-momentum in order for us to properly describe a conservation law for energy and momentum:  
\begin{equation}
\label{tmunu}
P^{\mu}=\frac{1}{c}\int_{S} (-g)(T^{\mu\nu}+t^{\mu\nu})dS_{\nu},
\end{equation}
where the quantity $t^{\mu\nu}$ is given by      
\begin{eqnarray}
\label{pseudo}
&&t^{\sigma\kappa}=\frac{c^4}{16 \pi G} [(2\Gamma^{\nu}_{\lambda\mu}\Gamma^{\theta}_{\nu\theta}-\Gamma^{\nu}_{\lambda\theta}\Gamma^{\theta}_{\mu\nu}-\Gamma^{\nu}_{\lambda\nu}\Gamma^{\theta}_{\mu\theta})(g^{\sigma\lambda}g^{\kappa\mu}-g^{\mu\kappa}g^{\lambda\mu})\nonumber\\&+&g^{\sigma\lambda}g^{\mu\nu}(\Gamma^{\kappa}_{\lambda\theta}\Gamma^{\theta}_{\mu\nu}+\Gamma^{\kappa}_{\mu\nu}\Gamma^{\theta}_{\lambda\theta}-\Gamma^{\kappa}_{\nu\theta}\Gamma^{\theta}_{\lambda\mu}-\Gamma^{\kappa}_{\lambda\mu}\Gamma^{\theta}_{\nu\theta})\nonumber\\&+&g^{\kappa\lambda}g^{\mu\nu}(\Gamma^{\kappa}_{\lambda \theta}\Gamma^{\theta}_{\mu\nu}+\Gamma^{\sigma}_{\mu\nu}\Gamma^{\theta}_{\lambda\theta}-\Gamma^{\sigma}_{\nu\theta}\Gamma^{\theta}_{\lambda\mu}-\Gamma^{\sigma}_{\lambda\mu}\Gamma^{\theta}_{\nu\theta})+g^{\kappa\lambda}g^{\mu\nu}\left(\Gamma^{\sigma}_{\lambda\mu}\Gamma^{\kappa}_{\mu\theta}-\Gamma^{\sigma}_{\lambda\mu}\Gamma^{\kappa}_{\nu\theta}\right)]. 
\end{eqnarray}

The expression above for $t^{\mu\nu}$ is achieved by means of a lengthy calculation which is provided step by step in the aforementioned reference Landau $\&$ Lifshitz~\cite{landau}, to which we refer the interested reader for full details. 
Given (\ref{tmunu}), the equivalent equation that indeed leads to a conservation law shall be
\begin{equation}
\label{divT1}
\partial_{\mu} \left[(-g)(T^{\mu\nu}+t^{\mu\nu})\right]=0.
\end{equation}

Notice that the nontensorial nature of $t^{\mu\nu}$ is evident due to its explicit dependence on products of Christoffel symbols. However, it does behave as a tensor under linear coordinate transformations, of which the Lorentz transformation is a particularly interesting case. In addition, notice that it vanishes in the locally inertial frame $x^{*}$, where $\Gamma^{\alpha}_{\mu\nu}(x^{*})=0$, at which the special relativity relation (\ref{4momentum}) is recovered. This enhances the fact that even the procedure leading to (\ref{divT1}) fails in providing a meaningful local conservation law that we could associate with the energy conservation, due to the absence in GR of a local meaning for the gravitational field. 

In the R. Wald textbook~\cite{wald}, the physical meaning of (\ref{ConsCurve}) is also discussed, although in a bit of a different way. The author shows that, in special relativity, the meaning of $\partial_{\mu}T^{\mu\nu}=0$ is unambiguous as a conservation expression for the energy--momentum tensor. To show this, let us consider a family of inertial observers with parallel worldlines, $v^{\alpha}$, which means 
\begin{equation}
\label{paral}
\partial_{\beta}v^{\alpha}=0.
\end{equation}

If we assume, for instance, a perfect fluid (although the reasoning actually holds for any matter and fields) described by $T^{\mu\nu}$, it is known that the quantity 
\begin{equation}
\label{current}    
J_{\alpha}=-T_{\alpha\beta}v^{\beta}
\end{equation}
shall correspond to the mass--energy current density $4-$vector as measured by these observers. Given (\ref{divT}), from (\ref{current}), it follows that  
\begin{equation}
\label{divJ}
\partial_{\alpha}J^{\alpha}=0.    
\end{equation}

The Gauss law tells us that the integration of (\ref{divJ}) over a four-dimensional volume, $V$, is equal to the surface integral below
\begin{equation}
\label{JnS}    
\int_{S} J^{\alpha}n_{\alpha}dS=0,    
\end{equation}
where $n_{\alpha}$ is the unit normal vector to $S$. The null flux defined by (\ref{JnS}) clearly indicates a conservation of energy, as it leads to the vanishing of the time variation of such a quantity inside the volume $V$. In other words, we can say that (\ref{divT}) is a requirement for the energy conservation as measured by a family of inertial observers.  

On the other hand, in a curved spacetime, as it is known, the relation (\ref{divT}) is replaced by (\ref{ConsCurve}). However, the presence of curvature spoils the interpretation of such a equation as a conservation law, since in this case there is no well-defined notion of parallel vectors at different points, which harms the introduction of a global family of inertial observers able to measure the energy of a distant particle. Let us see this feature in more detail. It is natural to think of the curvature-dependent extension of the condition (\ref{paral}) as
\begin{equation}
\label{paral1}
\nabla_{(\beta}v_{\alpha)}=0.
\end{equation}

Thus, the energy--momentum four vector (\ref{current}) now is the one measured by observers satisfying (\ref{paral1}). Nevertheless, in this case, the covariant derivative of $T^{\mu\nu}$ does not lead naturally to the covariant divergence of the current $J^{\mu}$, 
\begin{equation}
\label{divJ1}
\nabla^{\alpha}(T^{\mu\nu}v_{\nu})=0,
\end{equation}
which, by Gauss law, as we see before, this could ensure the conservation of energy. Because in the curved spacetime, in general, one is not able to define a family of observers satisfying $v^{\alpha}v_{\alpha}=-1$ and (\ref{paral1}). However, notice that an exception occurs if $v^{\mu}$ is a Killing vector that generates a one-parameter group of isometries in the spacetime. In this case, as it is well known, $v^{\mu}$ shall obey Equation (\ref{paral1}), which, in this context, is called the Killing equation. Thus, this inconsistency between (\ref{paral1}) and (\ref{divJ1}) prevents $\nabla_{\mu}T^{\mu\nu}=0$ to be interpreted as a requirement for a global energy conservation. In fact, let us recall that gravitational field itself, by means of tidal effects, can do work on a material system, thus altering locally its energy content. However, if we consider a small region of the space where such effects can be neglected, the energy of the system shall be conserved in a reasonable approximation. Thus, within such a small spacetime region, it is possible to define a vector field such that $\nabla_{\beta}v_{\alpha}\approx 0$. Thus, Equation (\ref{ConsCurve}) would indeed reflect an approximate conservation of energy as seen by these observers. Therefore, it is plausible to say that Ref. (\ref{ConsCurve}) represents a local conservation of the energy content of a physical system over small regions of spacetime.  

\section{Rastall Gravity}

The Rastall proposal extends GR causing a violation of the usual conservation law, making the covariant divergence of $T^{\mu\nu}$ proportional to the covariant  divergence of the  curvature scalar $R$ \cite{Rastall:1973nw}. While non-trivial to explain the nature of such a new source, this can be phenomenologically seen as the emergence of quantum effects in curved spacetimes as e.g., in the case of gravitational anomalies \cite{Bertlemann, BDavies}. This phenomenological approach and the absence of any variational formalism from which the field equations of Rastall theory could emerge have attracted the attention of many authors that tried to formulate a variational principle for the Rastall gravity. Some efforts in this sense can be found in the references \cite{smalley,Santos:2017nxm,DeMoraes:2019mef}.

In Rastall gravity, the conservation law is replaced by the equation
\begin{equation}
    T^{\mu\nu}_{\quad;\nu}=\frac{1-\lambda}{16 \pi G} g^{\mu\nu}R_{;\nu}\,
\end{equation}
where the GR framework is recovered by setting $\lambda= 1$. The associated field equations according to the Rastall’s proposal are
\begin{equation}
        R_{\mu\nu}-\frac{\lambda}{2}g_{\mu\nu}R={\kappa}T_{\mu\nu}.
\end{equation}

Within the context of Riemannian geometry, there is no variational principle associated with this theory, but similar structures may be found in the context of Weyl geometry~\cite{Almeida:2013dba}.

 Applications of the Rastall gravity to cosmology have been performed in Refs.~\cite{Fabris:2012hw, Batista:2011nu,Akarsu:2020yqa}. Black holes and other exact solutions have been studied in Refs.~\cite{Heydarzade:2016zof, Kumar:2017qws}. However, since Rastall theory should manifest mostly in high curvature environments, compact objects like neutron stars are perfect laboratories to constrain the parameter $\lambda$. In Ref.~\cite{Oliveira:2015lka}, by using realistic equations of state for neutron stars, interior conservative bounds on the non-GR behaviour of the Rastall theory have been placed at the 1\% level i.e., $\lambda < 0.01$.

 There is a long discussion on whether the Rastall theory (and also similar models \cite{Smalley}) is or is not equivalent to GR. Since the time this theory appeared in the 1970s, there have been claims that Rastall theory is an artificial construction of non-conserved quantities within a conservative theory \cite{LH}. This discussion has been revisited recently in Ref. \cite{Darabi:2017coc}. 

It is worth noting that there is a common criticism in the field of extended theories of gravity stating that any theory can be recast into the original GR form since the new geometrical terms appearing on the left-hand side of Einstein's equation can be sent to the right-hand side to assemble an effective energy--momentum tensor $T^{\mu\nu}_{eff}$. This is also the case of Rastall as discussed in~\cite{Visser:2017gpz}.

\section{Brans--Dicke Theory in the Einstein Frame}
It is well known that string theory predicts a scalar partner of the graviton in the low energy limits, the so-called dilatonic field (or dilaton). The Brans--Dicke theory is the simplest model in which such an extra degree of freedom shows up \cite{BD}. There are, however, two approaches by which such a theory can be studied \cite{faraoni1}. In the Jordan frame, the dilaton is a crucial piece of the geometrical sector in which it takes part as being co-responsible by the gravitational field, along with the metric tensor. However, in the Einstein frame, this scalar field is shifted to the matter sector, where it now shall couple to the ordinary matter. Both frames are related merely by a conformal transformation. As a consequence of this coupling, the paths followed by test particles become non-geodesic for a given spacetime and the standard covariant conservation of the energy--momentum deviates from the usual one.

In the Jordan frame, the gravitational action is

\begin{equation}
\label{JF}
S_{JF}=\frac{1}{16\pi} \int d^{4}x \sqrt{-g}\left[\phi R-\frac{\omega}{\phi}\nabla^{\mu}\phi\nabla_{\mu}\phi\right]+\int d^{4}x \sqrt{-g} {\cal L}_{m},  \end{equation}
where $\omega$ is the Brans--Dicke parameter and ${\cal L}_m$ is the Lagrangian of the matter fields. By extremizing the action above with respect to the metric, one has the following set of field equations:
\begin{eqnarray}
\label{feqBD}    
R_{\mu\nu}-\frac{1}{2}g_{\mu\nu}R=\frac{8 \pi}{\phi}T_{\mu\nu}+\frac{\omega}{\phi^2}\left(\nabla_{\mu}\phi \nabla_{\nu}\phi-\frac{1}{2}g_{\mu\nu}\nabla_{\alpha}\phi \nabla^{\alpha}\phi \right)+\frac{1}{\phi}\left(\nabla_{\mu}\nabla_{\nu}\phi-g_{\mu\nu} \Box \phi \right).  
\end{eqnarray}

The variation of the action with respect to the scalar field gives the dynamics obeyed by $\phi$
\begin{equation}
\label{boxPhi}    
\Box \phi = \frac{8 \pi T}{3+2\omega},    
\end{equation}
where $T$ is the trace of the energy--momentum tensor. Let us recall that the GR limit is achieved when $\omega \rightarrow{\infty}$ and $\phi\rightarrow{\phi_0=G^{-1}}$~\cite{weinberg}. In this frame, $T_{\mu\nu}$ conserves according to the standard condition:
\begin{equation}
\label{consJF}    
\nabla^{\nu}T_{\mu\nu}=0.
\end{equation}

It should be mentioned that there is an interesting version of the Brans--Dicke gravity in the Jordan frame in which the non-conservation of $T_{\mu\nu}$ shows up. In this alternative scenario, the Brans--Dicke model is combined with the Rastall theory, thus inheriting its non-conservative aspect~\cite{bdr}. This model was called Brans--Dicke-Rastall gravity. In~\cite{bdr}, the authors analyze some consequences of this theory both to the background cosmology and the parametrized post-Newtonian formalism.     

It is well known, however, that an alternative formulation for the Brans--Dicke theory is possible. It is achieved by means of the conformal transformation below
\begin{equation}
\label{confT}
g_{\mu\nu}\longrightarrow {\tilde{g}}_{\mu\nu}=G\phi g_{\mu\nu},    
\end{equation}
along with the following redefinition
\begin{equation}
\label{redPhi}
\phi \longrightarrow \tilde{\phi}=\int \sqrt{\frac{2\omega+3}{16\pi G}}\frac{d\phi}{\phi}. 
\end{equation}

Equations (\ref{confT}) and (\ref{redPhi}) lead the theory to the so-called Einstein frame. In this formulation, the gravitational action is rewritten as follows:
\begin{equation}
\label{EinsF}    
S_{EF}=\int d^4x \sqrt{-\tilde{g}}\left[\frac{\tilde{R}}{16\pi G}-\frac{1}{2}\tilde{g}^{\mu\nu}\tilde{\nabla}_{\mu}\tilde{\phi}\tilde{\nabla}_{\nu}\tilde{\phi}\right]+S_{M}\left(e^{-\alpha \tilde{\phi}}\tilde{g}_{\mu\nu},\psi\right),    
\end{equation}
where $\alpha \equiv \sqrt{\frac{16 \pi G}{2\omega+3}}$ and $\psi$ denotes the matter fields. Notice that, in this case, the redefined dilaton, $\tilde{\phi}$, couples minimally to the curvature, and the geometric sector is solely the Einstein--Hilbert action with $\tilde{\phi}$ acting as a matter field. This novel aspect arising in the Einstein frame reinterprets the role of the dilaton in the Brans--Dicke gravity, since now it indeed appears in the matter action, thus becoming able to couple with a given matter configuration. The main consequence of that is a departure of the energy--momentum conservation from its traditional expression. Now, this law is led to the following nonconservative form:
\begin{equation}
\label{consEF}    
\nabla^{\nu}T_{\mu\nu}=\alpha T \partial_{\mu}\phi.
\end{equation}

Notice that, leaving aside the trivial case of the GR limit (for which $\alpha\rightarrow 0$ and $\phi \rightarrow \phi_0$), the standard conservation can also occur for a given matter configuration whose energy--momentum tensor is trace free. 

\section{Gravity Theories from the Standard Model Extension}
In the gravity sector of the Standard Model Extension (SME), it is also possible to envisage physical properties with consequences to the energy--momentum conservation law. The SME is an effective theory encompassing the Standard Model of the particle physics and the General Relativity that incorporates possible deviations from the Lorentz and diffeomorphism symmetries~\cite{kost,kostA,kostB,kostC,kost1}. The breaking of diffeomorphism invariance is, in general, a remarkable feature of gravity theories arising from extensions of the standard model. Usually, this violation is induced by the presence of fixed background fields, which can break the local Lorentz and diffeomorphism symmetries either explicitly or spontaneously~\cite{bluhm}. In the former case, the background fields show up explicitly in the Lagrangian of the model, whereas, in the latter case, this does not happen, and the background just appears as a vacuum solution of the theory. It is well known, however, that, while the explicit breaking imposes difficulties for the gravity sector of the SME, with serious conflicts between the dynamical and the geometrical model’s constraints, the spontaneous one points towards a promising direction, free of such drawbacks. Nevertheless, as Kosteleck\'y shows in~\cite{kost}, these conflicts can be fixed in the context of Chern--Simons and massive gravity with an underlying Riemannian spacetime.
 
An effective gravity theory exhibiting Lorentz and diffeomorphism breaking can be generically represented by the action below (in the low energy limit)
\begin{equation}
\label{action}
S=S_{EH}+S_{LV}+S_{LI},    
\end{equation}
where $S_{EH}$ is the Einstein--Hilbert action,
\begin{equation}
\label{EH}
S_{EH}=\frac{1}{2\kappa}\int d^{4}x \sqrt{-g}R.    
\end{equation}

The Lorentz-violation term $S_{LV}$ is
\begin{equation}
\label{LV}
S_{LV}=\int d^{4}x \sqrt{-g}{\cal L}_{LV}(g_{\mu\nu},f^{\psi},\bar{k}_{\chi}),    
\end{equation}
where $f^{\psi}$ means the usual matter fields and $\bar{k}_{\chi}$ denoting the diffeomorphism violating background field. Moreover, $\psi$ and $\chi$ represent all the component indices of the tensors $f^{\psi}$ and $\bar{k}_{\chi}$, respectively.
In addition, finally, the Lorentz- and diffeomorphism-invariant term ${\cal L}_{LI}$ is given by
\begin{equation}
\label{LI}    
S_{LI}=\int d^{4}x \sqrt{-g}{\cal L}_{LI}(g_{\mu\nu},f^{\psi}).    
\end{equation}

As mentioned above, when the diffeomorphism symmetry breaking is explicit, the fixed background field $\bar{k}_{\chi}$ is nondynamical. This means that
\begin{equation}
\label{expl}    
(\delta S_{LV})_{\textrm{diff}}=\int d^{4}x \sqrt{-g}\frac{\delta {\cal L}_{LV}}{\delta \bar{k}_{\chi}}\delta \bar{k}_{\chi}\not= 0.    
\end{equation}

In addition, the variation of the full action (\ref{action}) gives
\begin{equation}
\label{eqs}    
G^{\mu\nu}=\kappa(T^{\mu\nu}_{LI}+T^{\mu\nu}_{LV}).    
\end{equation}

As clearly showed in Section \ref{sec:2}, the contracted Bianchi identity can be understood as a consequence of the diffeomorphism invariance of (\ref{EH}) in isolation. As this same invariance is not respected by $S_{LV}$, by taking the covariant divergence of (\ref{eqs}), one is left with 
\begin{equation}
\label{onshell}
\nabla_{\mu}(T_{LI}^{\mu\nu}+T_{LV}^{\mu\nu}) =0,
\end{equation}
which must hold on-shell. In this equation, the energy--momentum tensors $T_{LI}^{\mu\nu}$ and $T_{LV}^{\mu\nu}$ are defined in terms of ${\cal L}_{LI}$ and ${\cal L}_{LV}$, respectively, according to (\ref{emt}). Thus, in general, the conventional matter described by $T_{LI}^{\mu\nu}$ will not conserve as usual due to the presence of the inhomogeneous diffeomorphism breaking term in Equation (\ref{onshell}). However, it is usual to require the additional constraint $\nabla_{\mu}T_{LV}^{\mu\nu}=0$ in order to ensure the standard conservation condition~\cite{bluhm}. 

The equations above are presented in a quite general form, in order to cover arbitrary cases involving explicit spacetime symmetry breaking. However, following the spirit of the Ref.~\cite{bluhm}, we can look at specific examples. It is common that these models present conflicts between its dynamical description and some geometrical constraints. Such conflicts are usually evinced when the Bianchi identity is imposed onto the field equations. Let us analyze two examples that will help us to understand this aspect.
\subsection{Spacetime-Dependent Cosmological Constant}
For this model, the gravity is endowed with a spacetime-dependent cosmological constant. The total action of this theory is given by
\begin{equation}
\label{Lvar}
S=\int d^4 x \sqrt{-g} \left\{\frac{1}{2\kappa}[R-2\Lambda(x)]+{\cal L}_{m}\right\}.     
\end{equation}

By comparing (\ref{Lvar}) with (\ref{action}), we can set the mapping among the corresponding variables. Here, the fixed background is $\bar{k}_{\chi}=\Lambda(x)$, and the Lorentz-invariant Lagrangian is ${\cal L}_{LI}={\cal L}_{m}$, whereas the symmetry-violating piece is ${\cal L}_{LV}=-\Lambda(x)/\kappa$.  

Notice that, when $\Lambda(x)\not=0$ and is non-constant, the condition (\ref{expl}) applies, as the theory is endowed with a fixed, non-dynamical, background field given by $\Lambda(x)$ that breaks explicitly the diffeomorphism, since this field appears explicitly in the action. In fact, for this case, Equation (\ref{expl}) becomes
\begin{eqnarray}
\label{expLamb}    
(\delta S_{LV})_{\textrm{diff}}=\int d^{4}x \sqrt{-g}\frac{\delta {\cal L}_{LV}}{\delta \Lambda(x)}{\cal L}_{V}\Lambda(x)=-\int d^{4}x \sqrt{-g}\frac{1}{\kappa}V^{\mu}\partial_{\mu} \Lambda(x)\not=0,    
\end{eqnarray}
since the Lie derivative on $\Lambda(x)$ is ${\cal L}_{V}\Lambda(x)=V^{\mu}\partial_{\mu} \Lambda(x)$. 
The corresponding field equations are
\begin{equation}
\label{LvarEq}
G^{\mu\nu}=-\Lambda(x)g^{\mu\nu}+\kappa T^{\mu\nu}_{m},    
\end{equation}
where the symmetry-breaking energy--momentum tensor in the present example is $T^{\mu\nu}_{LV}=-g^{\mu\nu}\Lambda(x)/\kappa$. 
By using the contracted Bianchi identity on (\ref{LvarEq}), we obtain for this case the corresponding form for (\ref{onshell}), which is  
\begin{equation}
\label{Lcons}    
\nabla_{\mu}T^{\mu\nu}_{m}=(1/\kappa)g^{\mu\nu}\partial_{\mu}\Lambda(x).    
\end{equation}

Therefore, the presence of a non-constant $\Lambda(x)$ in the description of the gravity leads to a deadlock: if the standard conservation for $T^{\mu\nu}_{m}$ is required, Equation (\ref{Lcons}) tells us that necessarily $\partial_{\mu}\Lambda(x)=0$ implying in $\Lambda(x)=\textrm{const.}$, thus restoring the diffeomorphism invariance and contradicting the a priori assumption that $\Lambda(x)$ is non-constant. Notice that, in this case, the aforementioned conflict between the dynamics and the geometrical identities (in the present case, the Bianchi identity) is not evaded unless one assumes the non-trivial conservation law above (\ref{Lcons}). One may wonder if such a conflict is unavoidable, by raising the following question: is it possible to ensure the standard conservation law for $T^{\mu\nu}_{m}$ without necessarily restoring the diffeomorphism invariance? The next example provides an affirmative answer for this question.

\subsection{Chern--Simons Gravity} 
The so-called Chern--Simons term was first introduced in the context of three-dimensional gauge field theory and gravitational models \cite{CS3d,CS3dA}. Some years later, this same model was extended in order to represent a theory of gravity in the four-dimensional spacetime \cite{CSgravity1,CSgravity2}.
The action for the Chern--Simons gravity in four dimensions can be written as follows:
\begin{equation}
\label{CSaction}    
S_{CS}=\int d^{4}x\left[\frac{1}{2\kappa}\left(\sqrt{-g}R+\frac{1}{4}\theta^{\;*}RR\right)+\sqrt{-g}{\cal L}_{m}\right],    
\end{equation}
where $^{*}RR\equiv\;$${{^{*}R}^{\kappa}_{\;\lambda}}$${^{\mu\nu}R}^{\lambda}_{\;\;\kappa\mu\nu}$
is the gravitational Pontryagin density and ${{^{*}R}^{\kappa}_{\;\lambda}}^{\mu\nu}$$=\frac{1}{2}\epsilon^{\mu\nu\alpha\beta}$$R^{\kappa}_{\;\;\lambda\alpha\beta}$.
 The explicit breaking of diffeomorphism invariance occurs due to the embedding coordinate, $v^{\mu}$, which is related to the non-dynamical scalar $\theta(x)$ through $v^{\mu}\equiv\partial^{\mu}\theta$.   
The variation of the action (\ref{CSaction}) gives
\begin{equation}
\label{CSFeqs}    
G^{\mu\nu}+C^{\mu\nu}=\kappa T^{\mu\nu}_{m},    
\end{equation}
where $C^{\mu\nu}$ is the four-dimensional Cotton tensor which has the following form:
\begin{equation}
\label{cotton}
C^{\mu\nu}=-\frac{1}{2\sqrt{-g}}\left[v_{\sigma}\left(\epsilon^{\sigma\mu\alpha\beta}\nabla_{\alpha}R^{\nu}_{\;\;\beta}+\epsilon^{\sigma\nu\alpha\beta}\nabla_{\alpha}R^{\mu}_{\;\;\beta}\right)+\nabla_{\sigma}v_{\tau}\left(^{*}R^{\tau\mu\sigma\nu}+^{*}R^{\tau\nu\sigma\mu}\right)\right].    
\end{equation}

Due to the dependence of $C^{\mu\nu}$ upon the embedding coordinate $v^{\mu}$, the Cotton tensor encodes the information of diffeomorphism breaking in the field equations, so that we can set the following correspondence:
\begin{equation}
\label{cotton1}
T^{\mu\nu}_{LV}=\frac{C^{\mu\nu}}{\kappa}.
\end{equation}

By computing the covariant divergence of $C^{\mu\nu}$, we will have
\begin{equation}
\label{Dcotton}
\nabla_{\mu}C^{\mu\nu}=\frac{1}{8 \sqrt{-g}} (\partial^{\nu}\theta)^{\;*}RR.    
\end{equation}  
  
Looking at the action (\ref{CSaction}) and considering diffeomorphism transformations like (\ref{active}), Equation (\ref{expl}) shall be
\begin{equation}
\label{CSdiff}
(\delta S_{LV})_{\textrm{diff}}=\int d^{4}x \sqrt{-g}\frac{1}{4} (^{\;*}RR) V^{\mu}\partial_{\mu}\theta.    
\end{equation}

Thus, by the equation above, we find a twofold condition for the explicit breaking of diffeomorphism. The first condition is $\partial^{\mu}\theta\not=0$ and the second one is that the Pontryagin density is non-zero. We can examine the occurrence (or not) of the dynamics-geometry conflict by taking the covariant divergence of (\ref{CSFeqs}). If $\nabla_{\mu}T^{\mu\nu}_{m}=0$ is required, the Bianchi identity then obliges the relation below:
\begin{equation}
\label{divCotton}    
\nabla_{\mu}C^{\mu\nu}=0   
\end{equation}
to hold on shell. Since the divergence of $C^{\mu\nu}$ was also computed in (\ref{Dcotton}), the condition (\ref{divCotton}) imposes that $^{\;*}RR\;\partial_{\mu}\theta=0$. Thus, the conflict is evaded not only in the trivial situation when $\partial_{\mu}\theta=0$, implying in the restoration of the diffeomorphism invariance. It is evaded as well when the Pontryagin density vanishes, obeying the so-called Pontryagin constraint
\begin{equation}
\label{PontC}    
^{\;*}RR=0.    
\end{equation}

This is a constraint on the geometry that indicates a way to avoid the mentioned dynamics-geometry conflict by restricting the class of allowed geometries. It is known, for example, that any spacetime of Petrov types III, N, and O automatically satisfy (\ref{PontC}) (See~\cite{yunes}). 

On the other hand, if the usual conservation law $\nabla_{\mu}T^{\mu\nu}_{m}=0$ was somehow relaxed, when we took the covariant divergence of (\ref{CSFeqs}), we would have
\begin{equation*}
\label{NonCCS}    
\nabla_{\mu}T^{\mu\nu}_{m}=\frac{1}{8\kappa \sqrt{-g}}\partial^{\nu}\theta ^{\;*}RR,    
\end{equation*}
which means a deviation from the standard conservation law sourced by both the presence of $\partial^{\nu}\theta$ and the Pontryagin density.   

The formulation showed here for the Chern--Simons gravity is not the only one found in the literature. In \cite{yunes}, one can see an alternative one, where it is possible to render dynamics to the scalar field $\theta$. In this case, such a field shall obey a Klein--Gordon like equation of motion that is sourced both by stress-energy tensor and the curvature of spacetime. 

To end this section, let us mention an important attempt of setting experimental bounds on the non-dynamics Chern-Simons gravity. In \cite{smith}, by studying gravitomagnetic effects within this theory, the authors place important bounds on the parameter $m_{cs}$. In that work, they define such a parameter as $m_{cs}^{-1} \propto \dot{\theta}$ and assume the scalar field $\theta$ as being time varying but spatially homogeneous. They compute orbits of test bodies and the precession of gyroscopes in the linearized Chern--Simons gravity around a massive spinning bod. Then, they use observation from the LAGEOS \cite{lageos} and Gravity Probe B \cite{gpb} satellites to restrict $m_{cs}^{-1}$ to be less than 1000 km, which corresponds to $m_{cs} \geq 2\times 10^{-22}$ GeV.

\section{Emergent Gravity Theories Breaking General Covariance}

The so far weakly unexplored high energy limit of GR leaves room for investigation of emergent gravity theories i.e., approaches in which the low energy behavior appears as a manifestation of some yet unknown fundamental theory. 

Small violations of diffeomorphism invariance can be introduced into a physical theory in order to explore the  phenomenology behind emergent phenomena. As an example, one such approach has been discussed in Ref.~\cite{Anber:2009qp}. The general action proposed in this reference is of the form 
\begin{equation}
    \mathcal{L}=\frac{1}{2\kappa}\left[R+\sum_{i}a_i\mathcal{L}_i\right]+\mathcal{L}_m,
\end{equation}
where the $\mathcal{L}_i$ terms involve contributions that induce a violation of diffeomorphism invariance
% start a new page without indent 4.6cm
%\clearpage
\begin{eqnarray}
\mathcal{L}_1=-g^{\mu\nu}\Gamma^{\alpha}_{\mu\lambda} \Gamma^{\lambda}_{\nu\alpha}, \quad
&&\mathcal{L}_2=-g^{\mu\nu}\Gamma^{\alpha}_{\mu\nu} \Gamma^{\lambda}_{\lambda\alpha}, \quad
\mathcal{L}_3=-g^{\alpha\gamma}g^{\beta\rho}g_{\mu\nu}\Gamma^{\mu}_{\alpha\beta}\Gamma^{\nu}_{\gamma\rho}, \\ \nonumber
\mathcal{L}_4=-g^{\alpha\gamma}g_{\beta\lambda}g^{\mu\nu}\Gamma^{\lambda}_{\mu\nu} \Gamma^{\beta}_{\gamma\alpha}, \quad
\mathcal{L}_5&&=-g^{\alpha\beta}\Gamma^{\lambda}_{\lambda\alpha}\Gamma^{\mu}_{\mu\beta}, \quad
\mathcal{L}_6=-g^{\mu\nu}\partial_{\nu}\Gamma^{\lambda}_{\mu\lambda}, \quad
\mathcal{L}_7=-g^{\mu\nu}\partial_{\lambda}\Gamma^{\lambda}_{\mu\nu},
\end{eqnarray}
%\linenumbers
leading to the following field equations 
 \begin{equation}
 \label{diffEqs}
     R_{\mu\nu}-\frac{1}{2}g_{\mu\nu}R+a \mathcal{M}_{\mu\nu}=\kappa T_{\mu\nu}.
 \end{equation}
 
 The departure from GR is encoded in the new contribution $\mathcal{M}_{\mu\nu}$ defined as
 \begin{eqnarray}
     &&\mathcal{M}_{\mu\nu}=\mathcal{B}_{\mu\nu}+\mathcal{D}_{\mu\nu} \\ \nonumber
     &&\mathcal{B}_{\mu\nu}=-\frac{1}{2}g_{\mu\nu}g_{\alpha\beta}g^{\gamma\delta}g^{\epsilon\eta}\Gamma^{\alpha}_{\gamma\epsilon}\Gamma^{\beta}_{\delta \eta}+g^{\alpha\beta}g^{\gamma\delta}g_{\nu\phi}g_{\mu\epsilon}\Gamma^{\epsilon}_{\alpha\gamma}\Gamma^{\phi}_{\beta\delta}+2 g^{\phi\epsilon}g^{\alpha\gamma}g_{\delta\epsilon}g_{\phi\beta}\Gamma^{\beta}_{\mu\alpha}\Gamma^{\delta}_{\nu\gamma}\\ \nonumber
     &&\mathcal{D}_{\mu\nu}=\Gamma^{\lambda}_{\alpha\lambda}\mathcal{A}^{\alpha}_{\mu\nu}+\mathcal{A}^{\alpha}_{\mu\nu, \alpha},\\ \nonumber
     &&\mathcal{A}^{\alpha}_{\mu\nu}=g^{\alpha\beta}g_{\gamma\mu}\Gamma^{\gamma}_{\nu\beta}+g^{\alpha\beta}g_{\gamma\nu}\Gamma^{\gamma}_{\mu\beta}-\Gamma^{\alpha}_{\mu\nu}.
 \end{eqnarray}
 which is clearly not invariant under general coordinate transformation. 
 
 In order to implement a consistent condition on this set of equations in Ref.~\cite{Anber:2009qp}, it has been imposed the constraint below:
 \begin{equation}
     \label{constM}
 \nabla^{\mu}\mathcal{M}_{\mu\nu}=0
 \end{equation}
 
 In that reference, the authors analyzed this theory by expanding Equations (\ref{diffEqs}) along with the (\ref{constM}) in light of the recipe given by the PPN formalism. With this treatment, they aimed at constraining the diffeomorphism-breaking terms present in the model. As expected, this investigation shows that the parameters usually identified with the non-conservation of energy and momentum will not be zero for this model; they will in fact depend on the dimensionless parameter $a$ appearing in (\ref{diffEqs}). In addition, they found a strong bound on this parameter coming from the absence of preferred-frame effects in pulsars that leads $a$ to be less than $10^{-20}$ in gravitational strength.
 
\section{Action Dependent Lagrangian Theories}

This class of theory is based on the so-called Herglotz problem. The latter was originally built within a classical mechanics scenario, and consists of generalizing the action principle by introducing in the Lagrangian an action-dependence. Though non trivial, this kind of construction reads  
\begin{equation}
S= \int \mathcal{L}(x,\dot{x}, S)dt.
\end{equation}

This allows a proper description of dissipative phenomena in classical systems from first principles. Recently, Lazo {et al}~\cite{Lazo:2017udy}
 found that there exists a covariant generalization of this problem. Hence, a prototype of gravitational theory can be designed from  
\begin{equation}
\label{Lag}
\mathcal{L}=\sqrt{-g}(R-\lambda_{\alpha}s^{\alpha})+\mathcal{L}_m
\end{equation}
where quantity $s^{\alpha}$ is an action-density field. The coupling term $\lambda_{\mu}$ depends on the space-time coordinates. The interpretation employed in this approach refers to the action-dependence introduced in (\ref{Lag}) associated with $s^{\alpha}$ only with respect to the standard Einstein--Hilbert action. The matter action is not coupled to the $s^{\alpha}$ field. Therefore, departures from standard gravity provided by this theory are purely of geometric nature. As a result, this approach leads to a geometrical viscous gravity  model in which the dynamics of the theory is described by the generalized field equations 
\begin{equation}
\label{Keqs}
R_{\mu\nu}-\frac{1}{2}R g_{\mu\nu} + K_{\mu\nu}-\frac{1}{2}K g_{\mu\nu} = \kappa T_{\mu\nu}.
\end{equation}

 By applying the Bianchi identities to the above equation, one finds a relation involving $K_{\alpha\beta}$, its trace, and the matter sector. By considering a constant $G$ coupling, the system of field equations is sourced by the modified conservation law
\begin{equation}
 \label{consL}
 \kappa T^{\mu}_{\nu;\mu}=K^{\mu}_{\nu;\mu}-\frac{1}{2}K_{;\nu}.
\end{equation}
 
 The new aspect here is clearly encoded in the quantity $K_{\mu\nu}$ given by
\begin{equation}
K_{\mu\nu}= \lambda_{\alpha}\Gamma^{\alpha}_{\mu\nu}-\frac{1}{2}\left(\lambda_{\mu}\Gamma^{\alpha}_{\nu\alpha}+\lambda_{\nu}\Gamma^{\alpha}_{\mu \alpha}\right).
\end{equation}

The quantity $\lambda_{\alpha}$ plays the role of a background four-vector necessary in this nonconservative structure. 

In recent years, many gravitational problems have been investigated within this theory. In \cite{Fabris:2017msx}, the authors performed a study both of the FLRW background cosmology and the linear perturbative regime for this nonconservative gravity. They found that the background dynamics are equivalent to that one provided by the bulk viscous cosmology \cite{winfried}. On the other hand, the evolution of the linear perturbations indicated a possibility of avoiding, within the nonconservative theory, some of the problems present in the viscous scenarios \cite{velten1, velten2}. In \cite{Carames:2018atv}, the authors deepen such a cosmological study. In that work, they submitted the cosmology emerging from this theory to the scrutiny of some important cosmological datasets, both at the background and perturbative levels. This study revealed that the nonconservative cosmology was not viable, at least in the way it was originally formulated. However, the authors showed an interesting way out for this issue, by assuming that the matter conserves as usual, whereas the dark energy that obeys the non-standard conservation law becomes able to be pressureless. This new framework was revealed as a viable model in light of the analyzed cosmological data. In this study, we have computed the $f\sigma8$ observable for the perturbative regime of the theory. Next, we use the compilation of measurements of this quantity provided in \cite{rsd} to impose a stringent bound on the parameter of the theory, given by the interval $-0.9H_{0}<\lambda_{0}<-0.7H_{0}$, 
which also revealed compatible with $H(z)$ data, indicating a viable model both in the background and the perturbative levels.

This theory was also used to study cosmic string configurations, where were investigated the Abelian--Higgs strings as well as the phenomenological model of the Hiscock--Gott string, by means both of analytical and numerical techniques~\cite{CS}. The sum rules formalism for braneworld models within this nonconservative theory was examined in~\cite{brane}. In Ref.~\cite{Spher}, the authors discussed the conditions for the existence of static spherically symmetric solutions in this gravity (see also~\cite{Ayuso:2020vuu}).

\section{Nonminimal Curvature--Matter Coupling}

The principle of minimal coupling is evoked as one of the pillars to realize a gravitational theory. Alternative gravitational theories designed to stay close to GR maintain it. 
The consequences of abandoning this principle directly affect the way matter fields interact with geometry. If this principle is abandoned, the resulting field equations are non-trivial since the direct interplay between flat and curved spacetime, given by the familiar principle of General Covariance and the Equivalence Principle, is damaged. A direct consequence of adopting non-minimally couplings between the matter and geometric sectors is the appearance of nonconservative features.    
Extensions of $f(R)$ theories involving non-minimal couplings between curvature and the matter Lagrangian represent a class of theories in which $T^{\mu\nu}$ does not conserve. The family of $f(R, \mathcal{L}_m)$ theories is the typical prototype for this situation~\cite{Harko:2010mv} (see also~\cite{Obukhov:2013ona,tian}).
 
Particularly, the particle creation phenomena is a simple mechanism leading to the idea of non conservation. Ref.~\cite{Harko:2015pma} has discussed how one can associate this to a non-minimal coupling between the matter Lagrangian and curvature terms. This idea is designed by the following action:  
\begin{equation}
    S=\int \sqrt{-g} \left[\frac{ f_1(R)}{2\kappa}+f_2(R)\mathcal{L}_m\right],
\end{equation}
where $f_1(R)$ and $f_2(R)$ are arbitrary functions of the Ricci scalar. 

The energy--momentum tensor is defined in terms of the matter Lagrangian according to~(\ref{emt}). Then, one show that a general property of this type of theories is the non-conservation such that
\begin{equation}
    \nabla^{\mu}T_{\mu\nu}=\frac{\lambda F_2}{1+\lambda F_2}\left[g_{\mu\nu}\mathcal{L}_m-T_{\mu\nu}\right]\nabla^{\mu}R,
\end{equation}
where $F_i=f_{i,R}$ and $\lambda$ is a coupling constant that measures how strong the interaction is between $f_2(R)$ and the matter Lagrangian. Of course, the usual conservation law is recovered with $\lambda=0$.

This interpretation has been criticized, however, in Ref.~\cite{Azevedo:2019krx}. The reasoning of the criticism contained in the latter reference is that the non-minimal coupling actually induces a change in the particle--momentum on a cosmological timescale, which can not be associated with the particle creation process.

Observational constraints on this class of theories can be found in, e.g., Ref.~\cite{An:2015mvw}.

\section{\boldmath{$f(R,T)$} Theories}

The non-minimal coupling between matter and curvature terms has been indeed widely studied.
One such proposal is the so-called $f(R,T)$ theory where $T = T^{\mu}_{\mu}$ is the trace of the stress--energy tensor. Of course, the GR limit of such theory corresponds to $f(R,T)=R$ in the action
\begin{equation}
S=S_G+S_m=\frac{1}{2 \kappa}\int d^4 x \sqrt{-g} f(R,T)+\int d^4 x \sqrt{-g} \mathcal{L}_m
\end{equation}

%where $\kappa=8\pi G$ and $\mathcal{L}_m$ is the Lagrangian density of the matter fields.
This approach has been introduced in Ref.~\cite{Harko:2011kv} aiming to describe running cosmological constant cosmologies. Therefore, this modification of gravity trying to explain the accelerated expansion of the universe and dark matter seems to be a fundamental ingredient in the viable $f(R,T)$ scenarios~\cite{Velten:2017hhf}.

By defining $T_{\mu\nu}$ as \footnote{With a minus sign with respect to (\ref{emt}).}
\begin{equation}
\label{emt1} 
T_{\mu\nu}=g_{\mu\nu}{\cal L}_m-2\frac{\partial {\cal L}_m}{\partial g^{\mu\nu}}.
\end{equation}

By varying the action with respect to the metric (as in the standard metric formalism),
\begin{eqnarray}
\label{feqs}
f_R(R,T)R_{\mu\nu}-\frac{1}{2}f(R,T)g_{\mu\nu}+(g_{\mu\nu} \Box-\nabla_{\mu}\nabla_{\nu})f_R \left(R,T\right) = \left[\kappa -f_T(R,T)\right]T_{\mu\nu}-f_T \Theta_{\mu\nu},
\end{eqnarray}
where
\begin{equation}
\label{theta}
\Theta_{\mu\nu}\equiv g^{\alpha \beta} \frac{\delta T_{\alpha\beta}}{\delta g^{\mu\nu}}=-2T_{\mu\nu}+g_{\mu\nu}\mathcal{L}_m-2 g^{\alpha\beta}\frac{\partial^2 \mathcal{L}_m}{\partial g^{\mu\nu}\partial g^{\alpha\beta}}.
\end{equation}

In addition, the conservation law to be obeyed by $T_{\mu\nu}$ in this case shall be
\begin{equation}
\nabla^{\mu}\left[(\kappa-f_{T})T^{\mu\nu}-f_{T}\Theta_{\mu\nu}\right]=0.    
\end{equation}

This condition is easily obtained by taking the covariant divergence of (\ref{feqs}), bearing in mind that the left-hand side of these equations has null divergence, as can be straightforwardly verified (see~\cite{koivisto}).

In a FLRW background, an energy--momentum tensor of a perfect fluid is sourced written in terms of 
%Please confirm this change
its energy density $\rho$ and the pressure $p$. From the above definitions, we can write the Lagrangian as $\mathcal{L}_m=-p$, while the tensor (\ref{theta}) reduces to $\Theta_{\mu\nu}=-2T_{\mu\nu}-pg_{\mu\nu}$. Let us focus on a class of $f(R,T)$ theories given by $f(R,T)=f_1(R)+f_2(T)$. The background expansion obeys equations
\begin{equation}
\label{eq00}
-3(\dot{H}+H^2)f'_{1}-\frac{f_{1}}{2}-\frac{f_{2}}{2}+3H\dot{f'_{1}}=\kappa \rho+f'_{2}(1+w)
\end{equation}
and
\begin{equation}
\label{eq11}
(\dot{H}+3H^2)f'_{1}+\frac{f_{1}}{2}+\frac{f_{2}}{2}-\ddot{f'_{1}}-2H\dot{f'_{1}}=\kappa p, 
\end{equation}
where the prime and dot denote derivatives with respect to the argument, i.e., $f^{\prime}_1= d\, f_1(R) / dR$, and to the cosmic time, respectively. 

In practice, any$f(R,T)$ model gives rise to a deviation from the usual conservation law such that 
\begin{equation}
\label{consfrt}
\dot{\rho} +3 H\rho(1+w)= -\frac{1}{\kappa+f'_{2}}\left[(1+w)\rho \dot{f'_{2}}+ w \dot{\rho} f'_{2}+\frac{1}{2}\dot{f}_{2}\right],
\end{equation}
where $p=w\rho$. Notice that (\ref{eq00})--(\ref{consfrt}) form a system of three independent differential equations. The fourth-order derivatives of the metric appearing in these equations, $\ddot{f'_{1}}$, gives rise to new degrees of freedom in $f(R,T)$ theory, so that, along with the variables $a$ and $\rho$, it is also necessary to consider $\ddot{a}$ as an independent variable in order to provide a solution for such system of equations.

%The covariant conservation of $T_{\mu\nu}$ is an essential feature of GR, which manifests as an immediate consequence of the diffeomorphism invariance of the theory. Thus,  it is expected that any classical gravitational theory shall in principle obey such requirement as well. 

In the context of interacting dark energy models, a local violation of $\nabla_{\mu}T^{\mu\nu}=0$ may be allowed by means of a possible exchange of either energy or momentum (or both) between the two dark components. Nonetheless, even in these models, this exchange occurs in such a way as to preserve the conservation of the total dark fluid. Differently, Equation (\ref{consfrt}) shows a non-conservation of the matter--energy content as a whole, revealing a significant drawback of this class of $f(R,T)$ theories. 
In light of this, in~\cite{alvarenga}, the authors imposed by hand the fulfillment of (\ref{consfrt}) by setting to zero the expression inside the bracket, e.g $(1+w)\rho \dot{f'_{2}}+ w \dot{\rho} f'_{2}+\frac{1}{2}\dot{f}_{2}=0$. By using a chain rule, one can get rid of the time derivatives and write this constraint condition as a second order differential equation for the function $f_2(T)$, whose integration provides a solution in the form
\begin{equation}
\label{sol}
f_2(T)=\sigma T^{\frac{3w+1}{2(w+1)}}+\sigma_0,
\end{equation}
where $\sigma$ and $\sigma_0$ are integration constants. We may avoid the trivial case $f_2(T)=\textrm{const.}$ by assuming the necessary condition $\omega\neq - 1/3$. In addition,  $\omega\neq + 1/3$ is adopted in order to assure that $T \neq 0$. Considering pressureless matter, $\omega=0$, the solution becomes
\begin{equation}
\label{sol}
f_2(T)=\sigma T^{\frac{1}{2}}+\sigma_0.
\end{equation}

Then, this is the only choice assuring conservation, as it constitutes the only case in which the standard conservation law is preserved, which implies automatically ruling out anyone else if one demands that conservation is required. It is not surprising that the usual conservation condition is gained in ``separable'' $f(R,T)$ models, namely those ones obeying $f(R,T)=f_1(R)+f_2(T)$. Let us recall that the departure from the traditional conservation law in arbitrary $f(R,T)$ theories arises precisely because of the non-minimal coupling between curvature and matter. When these two sectors appear in such a separated added up terms as $f(R,T)=f_1(R)+f_2(T)$, it makes it possible to obtain a differential equation only for $f_2(T)$, decoupled from the function $f_1(R)$, therefore free from any $R$-dependence. As we have seen in (\ref{consfrt}), this differential equation constitutes a constraint leading to the usual conservation law. Thus,  it is somehow expected that, when a non-coupling is imposed, the standard conservation appears as a particular case. It is also possible to redesign the $f(R,T)$ theory in such a way that it evades the continuity equation by adding the extra geometric terms to sum up the effective energy--momentum tensor~\cite{Moraes:2016mlp}.

However, if one intends to assume the usual conservation, a stringent restriction on $f(R,T)$ gravity applies. By taking this path, Refs.~\cite{alvarenga} as well as 
\cite{Velten:2017hhf} bring the message that versions of $f(R,T)$ models based on the separation $f(R,T)=f_{1}(R)+f_{2}(T)$ are disfavored in light of recent data and therefore cannot be interpreted as viable theories. Furthermore, there is a more profound argument to not take into consideration this class of $f(R,T)$ models. In a recent discussion, it was claimed that the term $f_2(T)$ may be simply incorporated into the matter Lagrangian $L_m$, which means that it is not possible to physically separate their effects as they depend on the same variables and are added up in the total action~\cite{fisher1}. Thus,  these models would consist of a mere redefinition of the matter sector without bringing any new information or physical effect to the problem under study. Other authors, however, have disputed this claim. We direct the reader to Refs.~\cite{harko1, fisher1}, where it is possible to follow the entire debate on it. 

The discussion on the conservation properties in $f(R, T^\phi)$, a scalar field variant formulation, has been recently discussed in~\cite{Singh:2018tlm}.

\section{Nonconservative Traceless Gravity}

Unimodular gravity is a well-known alternative gravitational theory. In this approach, the cosmological constant appears naturally in the form of an integration constant. By obtaining the field equations from the Einstein--Hilbert Lagrangian imposing the condition $g_{\mu\nu}\delta g^{\mu\nu}=0$, one finds
\begin{equation}
    G_{\mu\nu}+\frac{1}{4}g_{\mu\nu}R=\kappa \left( T_{\mu\nu}-\frac{1}{4}g_{\mu\nu}T\right).
\end{equation}

In addition, by taking the divergence of the above equation and using the Bianchi identities,
\begin{equation}\label{unimodular2}
    \frac{R^{,\nu}}{4}=8\pi G \left( T^{\mu\nu}_{\quad; \mu}-\frac{T^{;\nu}}{4}\right).
\end{equation}

It is worth noting that there is an extra constraint on the determinant of the metric, and therefore there are nine independent equations of motion, one less than GR. Then, energy--momentum conservation ad hoc imposed in this theory.

Using the approach adopted in Ref.~\cite{Gao}, it is possible to design a non-conservation unimodular theory. In Ref.~\cite{Daouda:2018kuo}, a constant curvature $R=$const. has been used to constrain unimodular gravity. If this condition is applied to Equation (\ref{unimodular2}), one immediately finds
\begin{equation}
    T^{\mu\nu}_{\quad}=\frac{T^{;\nu}}{4}.
\end{equation}

A consequence of this case is the fixing of the scaling law for the enthalpy of the system such that
\begin{eqnarray}
\rho + p = C a^{-4},
\end{eqnarray}
where $C$ is a constant.

It is also noted in Ref.~\cite{Daouda:2018kuo} that other constraining conditions such as e.g., when the quantity $\sqrt{-g}(R + 4 \Lambda)$ is constant, also lead to nonconservative models with background expansions similar to the $\Lambda$CDM, but with distinct perturbative behavior. 

\section{Energy Conditions When \boldmath{$T^{\mu\nu}_{ \quad; \nu} \neq 0$}.}\label{sec:energy}

The interface between non-conservation in modified gravity theories and how energy conditions are employed is worth being highlighted. Most of the theories discussed in this work are considered nonconservative since new geometric terms appear on the left-hand side of their field equations. Then, very often seen in the literature, in order to maintain the application of the contracted Bianchi identities to the Einstein tensor $G_{\mu\nu}$, all remaining new geometric contributions are sent to the right-hand side of field equations to compose an effective $T^{eff}_{\mu\nu}=\bar{T}_{\mu\nu}$. While formally possible to impose in an ad hoc manner that such new $\bar{T}_{\mu\nu}$ conserves i.e., $\bar{T}^{\mu\nu}_{\, ; \mu}=0$, this procedure leads to a different interpretation e.g., on how the familiar energy conditions should apply to the $\bar{T}_{\mu\nu}$ components.

Let us sketch now the basic idea behind the above argument. At the field equations level, one can formally recast most of the modified gravity theories in the form \begin{equation}\label{ETG}
\sigma(\Psi^i)\left(G_{\mu\nu} + W_{\mu\nu}\right)= \kappa T_{\mu\nu},
\end{equation}
where the factor $\sigma(\Psi^i)$ is a coupling to the gravity while $\Psi^{i}$ represents curvature invariants or other fields, like scalar ones. The symmetric tensor $W_{\mu\nu}$ stands for additional geometrical terms which may appear in specific theory under consideration. We want to mention that, in Equation (\ref{ETG}), the energy--momentum tensor $T_{\mu\nu}$ will be considered as the one of a perfect fluid as defined in (\ref{EM}). Of course, Equation (\ref{ETG}) does not encompass all the possible alternatives to GR at the field equations level. However, most of the main proposals in the market (including most of the theories discussed in this work) can be reshaped in this form. From the structure presented in (\ref{ETG}), one identifies that GR is immediately recovered if $\sigma(\Psi^i)=1$ and $W_{\mu\nu}=0$. Equation (\ref{ETG}) can also be rewritten as a GR-like theory according to
\begin{equation}\label{EffEqETG}
    G_{\mu\nu}=\kappa \bar{T}_{\mu\nu}=\frac{\kappa}{\sigma}T_{\mu\nu}-W_{\mu\nu}.
\end{equation}

In principle, one can not postulate that the effective energy--momentum tensor is conserved i.e., $\bar{T}^{\mu\nu}{\,;\mu=0}$. However, this is indirectly achieved due to the Bianchi identities. Then, what is the meaning of the former statement ($\bar{T}^{\mu\nu}_{\, ;\mu=0}$)?

This issue has been extensively discussed in Refs.~\cite{Capozziello:2013vna,Capozziello:2014bqa} (see also~\cite{Visser:1999de}), and we briefly review it now since this topic is critical for the discussion of the viability of modified gravity models.

Firstly, let diffeomorphism invariance take place in the perfect fluid matter action i.e., $T^{\mu\nu}_{\,;\mu}=0$. On the other hand, contracted Bianchi identities assure that $G^{\mu\nu}_{\,;\mu}=0$. Then, if both conditions take place, there appears a new constraint equation derived from (\ref{EffEqETG}). It reads
\begin{equation}
    W^{\mu\nu}_{\quad; \mu}=-\frac{\kappa}{\sigma^2}T^{\mu\nu}\sigma_{\,; \mu}.
\end{equation}

However, if one evades the diffeomorphism invariance of the matter action, the physical meaning of the above constraint can be translated to the notion of a non-conservation, i.e., the mere application of the Bianchi identities to the field Equation (\ref{EffEqETG}) yields to
\begin{equation}
    T^{\mu\nu}_{\quad;\nu}=\left(\sigma W^{\mu\nu}\right)_{;\,\nu}+\left(\frac{\sigma_{;\,\nu}}{\sigma}\right)\left[T^{\mu\nu}-\left(\sigma W^{\mu\nu}\right)\right],
\end{equation}
implying a different geodesic structure for the matter fields.

As an example of how delicate this issue is, let us focus on the weak energy condition as a simple instance of a situation in which care must be taken when comparing the structure of modified gravity theories with GR. In GR, the weak energy condition is usually interpreted as the one that guarantees positiveness of energy density in a locally inertial frame i.e., $\rho>0$. Actually, in a coordinate-independent way, the weak energy condition reads
\begin{equation}
    T_{\mu\nu}\,U^{\mu}\,U^{\nu}\geqslant 0,
\end{equation}
where $U^{\mu}$ is any timelike vector. 

It is a common practice in the literature to replace $T_{\mu\nu}$ in the above inequality by $\bar{T}_{\mu\nu}$ i.e., one writes down an effective energy density $\bar{\rho}$ which depends on the curvature terms and interprets $\bar{\rho}\geqslant 0$ as the true energy condition in modified gravity theories. As shown in Ref.~\cite{Capozziello:2014bqa}, this is not exactly true, not only for the weak energy conditions but also for the remaining ones, and care must be taken when analyzing energy conditions in modifications of GR, in particular, proposals in which the conservation of the energy momentum energy is not observed. As a simple example, for a given geometric contribution $W_{00}$, one can find situations unacceptable situations where $\bar{\rho}\geqslant 0$ while the physical $\rho$, derived from the microphysical description of the matter fields, is negative.

\section{Conclusions}

Modified gravity theories represent a fruitful way to study the astronomical observations behind the dark matter/energy phenomena. While the direct searches for the cosmic dark components do not deliver positive results proving the existence of exotic matter fields in nature, there is room for investigations aiming to explain the observed astrophysical/cosmological data via the introduction of geometrical features beyond GR. In order to promote departures from GR, one has to either abandon one (or more) of the pillars over which GR has been built up, or add new fields. A modification that leads to a new modified gravitational theory can be seen as more or less radical depending on how strong the assumptions it is based on are.   

In this contribution, we have discussed some of the attempts found in literature to explore alternative gravity theories in which the null covariant divergence of the energy--momentum tensor is not achieved.  Such non-conservation appears in a modified gravitational theory by different methods. As reviewed in this work, this occurs either by imposing in an ad hoc manner the non-vanishing of the covariant derivative of the energy--momentum tensor or obtaining this feature as a consequence from a first principle construction. 
Though we have probably failed to mention all existing proposals, the main ideas that have motivated current research in this field have been discussed here. In addition,  interesting proposals in which there appears violation of the energy--momentum conservation include e.g., cosmological diffusion effects~\cite{Calogero:2013zba,Josset:2016vrq,Perez:2020cwa} and other physical mechanisms~\cite{Banerjee:2008aa,Maulana:2019sgd,Lobato:2018vpq,Pan:2016jli}.

We have also revisited in Section \ref{sec:energy} how subtle the direct application is of well-established results of the general relativistic framework to the case of nonconservative theories. In particular, the application of energy conditions when $T^{\mu\nu}_{\quad;\nu}\neq 0$ requires a careful treatment not usually seen in the literature.    

While one can not find conclusive evidence that nonconservative theories of gravity should be ruled out as viable alternatives, they will stay on the market and be a matter of intense investigation as seen currently in the literature.

%%%%%%%%%%%%%%%%%%%%%%%%%%%%%%%%%%%%%%%%%%
\vspace{6pt}

%%%%%%%%%%%%%%%%%%%%%%%%%%%%%%%%%%%%%%%%%%
\acknowledgments{We thank three anonymous reviewers for provided helpful comments on earlier drafts of the manuscript. We acknowledge discussions with Jose Beltrán Jimenez, Júlio Fabris, Oliver Piattella and Saulo Carneiro. This research was partially funded by CAPES, CNPq, FAPES and Proppi/UFOP.}

%%%%%%%%%%%%%%%%%%%%%%%%%%%%%%%%%%%%%%%%%%
\end{document}